# Releasing H$_2$ molecules with a partial pressure difference without the use of temperature

Hoonkyung Lee,[1,*] Bing Huang,[2] Wenhui Duan,[2] and Jisoon Ihm[1]

[1]Department of Physics and Astronomy, Seoul National University, Seoul, 151-747, Korea

[2]Department of Physics, Tsinghua University, Beijing 100084, People's Republic of China

[*]Corresponding author: hkiee3@snu.ac.kr

## ABSTRACT

Using the pseudopotential density functional method as well as equilibrium thermodynamic functions, we explore the process of releasing H$_2$ molecules adsorbed on a transition metal atom caused by the hydrogen-ammonia partial pressure difference. The H$_2$ molecules bind to a transition metal atom at H$_2$ pressure-NH$_3$ pressure-temperature 50 atm-10$^{-9}$ atm-25 °C, and they are released at 3 atm-10$^{-6}$ atm-25 °C. This process involves the same mechanism responsible for carbon monoxide poisoning of hemoglobin with the O$_2$-CO partial pressure difference. We show that our findings can be applicable to an approach to induce hydrogen desorption on nanostructured hydrogen storage materials without the need for increasing temperature.

# I. INTRODUCTION

Transition metal (TM) atoms together with $H_2$ molecules form a TM-$H_2$ complex caused in part by the so-called Kubas interaction,[1,2] which has an unusual bonding character distinct from conventional bondings such as covalent, ionic, metallic, or van der Waals bondings. This bonding configuration offers a rare opportunity for the development of reversible hydrogen storage media for near room temperature operation[3-15] because the binding energy of the $H_2$ molecule in the system is theoretically estimated to be in the range of 0.2−0.6 eV[16]. Recently, an enhanced interaction of $H_2$ molecules on organic Ti complexes with a binding energy of ~0.2 eV has been observed[10]. On the other hand, temperature variation methods have typically been used for desorption, i.e., for delivering hydrogen from the storage tank. However, it has turned out that TM-decorated nanostructured materials which have a binding energy of 0.2−0.6 eV[3-8] might result in a high desorption temperature of ~200 °C because the binding energy of $H_2$ molecule should be in the narrow window of 0.25−0.35 eV to release the $H_2$ by a temperature of ~100 °C[5]. This implies that, like metal or chemical hydrides, nanostructured materials may result in a high desorption temperature even in the molecular hydrogen storage form. This has practical importance since increasing high temperature on vehicles causes inevitable high energy consumption and is an obstacle to practical applications.

Here, we make the suggestion of releasing $H_2$ molecules without a temperature change in analogy to carbon monoxide poisoning of hemoglobin (Hb)[17]. When Hb is exposed to a mixed gas of oxygen and carbon monoxide, the occupation of $O_2$ or CO molecules on Hb is determined by the competing adsorption probability of $O_2$ and CO, i.e., the Gibbs factor ( $e^{(\mu_i - \varepsilon_i)/kT}$ ) where $\mu_i$ (<0) and $-\varepsilon_i$ (>0) for $i$=$O_2$ or CO are the chemical potential of the gas and the binding energy of the molecules on Hb, respectively. When the partial pressure of $O_2$ ( $p_{O_2}$ ) and CO ( $p_{CO}$ ) gas satisfies the relation, $p_{O_2} \gg p_{CO}$ ( $\mu_{O_2} \gg \mu_{CO}$ ), $O_2$ molecules bind to Hb as displayed in Fig. 1(a) because the Gibbs factor for oxygen binding dominates ( $e^{(\mu_{O_2} - \varepsilon_{O_2})/kT} \gg e^{(\mu_{CO} - \varepsilon_{CO})/kT}$ ). When the pressure of CO gas approaches that of $O_2$ gas $p_{O_2} > p_{CO}$ ($\mu_{O_2} > \mu_{CO}$), the $O_2$ molecules are released and CO molecules bind to Hb because the Gibbs factor for CO binding dominates ($e^{(\mu_{O_2} - \varepsilon_{O_2})/kT} << e^{(\mu_{CO} - \varepsilon_{CO})/kT}$). This is attributed to a binding energy of the CO molecule (~0.7 eV) which is larger than that of the $O_2$ molecule (~0.5 eV)[18]. This shows that the CO molecule corresponds to an "$O_2$-releasing molecule", and adsorption of $O_2$ molecules on Hb can be controlled by the $O_2$-CO partial pressure difference. In this paper, we find a "$H_2$-releasing molecule", $NH_3$ molecule which has approximately twice as large as a binding energy of $H_2$ molecules on a TM atom. Employing the $H_2$-$NH_3$ partial pressure difference, we will show that releasing $H_2$ molecules adsorbed on a TM atom as illustrated in Fig. 1(b) and its application to a hydrogen desorption in TM-decorated hydrogen-storage nanomaterials.

## II. COMPUTATIL DETAILS

Our calculations were carried out using the pseudopotential density functional method with a plane-wave-based total energy minimization[19] within the generalized gradient approximation (GGA)[20], and the kinetic energy cutoff was taken to be 476 eV. The optimized atomic positions were obtained by relaxation until the Hellmann-Feynman force on each atom was less than 0.01 eV/Å. Supercell[21] calculations were employed throughout where atoms between adjacent nanostructures were separated by over 10 Å.

## III. RESULTS AND DISCUSSION

To show the feasibility of this idea explicitly, we choose recently-studied Sc-, Ti-, and V-decorated *cis*-polyacetylene (cPA) and ethane-1,2-diol (ETD, $C_2H_6O_2$) as a hydrogen storage medium.[5,22] We calculate the binding energy of $H_2$ and $NH_3$ molecules as a function of the number of adsorbed $H_2$ and $NH_3$'s on the TM atom. Multiple $H_2$ and $NH_3$ molecules both bind to a TM atom. Figure 2 shows some optimized geometries for the configuration with adsorbed $H_2$ and $NH_3$ molecules, and the calculated (static) binding energy of the $H_2$ and $NH_3$ per the number of adsorbed $H_2$ and $NH_3$'s. The distance between the $H_2$ ($NH_3$) molecule and the TM atom is ~1.9 (~2.2) Å and the bond length of the $H_2$ molecules is slightly elongated to ~0.83 Å from 0.75 Å for an isolated molecule. According to reference,[23] TM-$H_2$ complexes are observed, and the hybridization of the TM $d$ states with the $NH_3$ states as well as the induced polarization of the $NH_3$ molecule both contribute to the $NH_3$ binding.

To describe a thermodynamic situation of multiple binding of both $H_2$ and $\chi$ molecules to a site (TM) where $\chi$ indicates a $H_2$-releasing molecule (i.e., $NH_3$ molecule), we obtain the occupation numbers for $H_2$ and $\chi$ molecules per site as a function of the $H_2$-pressure and $\chi$-pressure and temperature. For equilibrium conditions between the hydrogen ($H_2$-releasing) gas and the adsorbed $H_2$ ($\chi$) where the chemical potentials $\mu_{H_2}$ and $\mu_\chi$ of the surrounding hydrogen and $H_2$-releasing gas describe the thermal and particle reservoir, the grand partition function is given by

$$Z = \sum_{n=0}^{N_{max}^{H_2}} \sum_{m=0}^{N_{max}^{\chi}} g_{nm} e^{(n\mu_{H_2} + m\mu_\chi - (n+m)\varepsilon_{nm})/kT}$$

for a site where $-\varepsilon_{nm}$ (>0) is the binding energy of adsorbed $H_2$ and $\chi$ per the number of $H_2$ and $\chi$ molecules when the number of the adsorbed $H_2$ and $\chi$ molecules per site is $n$ and $m$, respectively, and $k$ and $T$ are the Boltzmann constant and the temperature, respectively. The maximum number of adsorbed $H_2$ ($\chi$)'s per site is $N_{max}^{H_2}$ ($N_{max}^{\chi}$), and $g_{nm}$ is the degeneracy of the configuration for a given $n$ and $m$. The fractional occupation number $f_i$ per site where $i$ indicates a kind of gas (i.e., $i = H_2$ or $\chi$) is obtained from the relation of $f_i = kT\partial \log Z/\partial \mu_i$ :

$$f_{H_2(\chi)} = \frac{\displaystyle\sum_{n=0}^{N_{max}^{H_2}} \sum_{m=0}^{N_{max}^{\chi}} g_{nm} n(m) e^{(n\mu_{H_2} + m\mu_\chi - (n+m)\varepsilon_{nm})/kT}}{\displaystyle\sum_{n=0}^{N_{max}^{H_2}} \sum_{m=0}^{N_{max}^{\chi}} g_{nm} e^{(n\mu_{H_2} + m\mu_\chi - (n+m)\varepsilon_{nm})/kT}} \tag{1}.$$

This formula can also be applicable for the situation of $O_2$ and $CO$ binding to hemoglobin (i.e., $CO$ poisoning of hemoglobin) as well as the adsorption of $O_2$ ($H_2$) molecules on Hb (TM).[5]

We next explore the thermodynamics for the adsorption of $H_2$ molecules on a TM atom in the absence of ammonia gas. Here we consider the chemical potentials for $H_2$ and $NH_3$ gases as in Ref. 24, and the binding energy of the $H_2$ ($NH_3$) molecules is subtracted by 25 (3)% from the calculated (static) binding energy presented in Fig. 2(g) because of the zero-point vibration energy[5]. Employing Eq. (1) with the actual binding energy $-\varepsilon_{n0}$, we calculate the occupation number of $H_2$ molecules as a function of the pressure and the temperature for the Ti-decorated cPA as shown in Fig. 3(a). Five $H_2$ molecules adsorb on the Ti atom at 50 atm and 25 ºC because the Gibbs factor ($e^{5(\mu_{H_2} - \varepsilon_{50})/kT}$) for the adsorption of 5 $H_2$ molecules dominates ($\varepsilon_{50} = -0.38$ eV, $\mu_{H_2} = -0.21$ eV). About two $H_2$ molecules are not desorbed at 3 atm and 80 ºC because the term $e^{2(\mu_{H_2} - \varepsilon_{20})/kT}$ dominates ($\varepsilon_{20} = -0.41$ eV, $\mu_{H_2} = -0.36$ eV). However, a temperature as high as 200 ºC is necessary to fully release the $H_2$ molecules because $\mu_{H_2} < \varepsilon_{n0}$ ($\mu_{H_2} = -0.52$ eV) at 3 atm and 200 ºC. Since the chemical potential difference between the desorption (50 atm and 25 ºC) and the adsorption (3 atm and 80 ºC) conditions based on the goal of the U.S. Department of Energy (DOE)[25] is $-0.21\sim-0.36$ eV as shown in Fig. 3(b), TM-decorated nanostructured hydrogen-storage materials discussed in the literature[3-7] may be inadequate to meet the DOE goal, and a high desorption temperature (>200 ºC) is needed.

We next investigate the effects of ammonia gas on adsorption of $H_2$ molecules for the Ti-decorated cPA. Using Eq. (1) with energy $-\varepsilon_{nm}$, the occupation number of $H_2$ ($NH_3$) molecules as a function of the partial pressures of hydrogen and ammonia gases at 25 °C is calculated. Five $H_2$ molecules adsorb on the Ti atom at $H_2$ pressure-$NH_3$ pressure 50 atm-$10^{-9}$ atm and they are released by the adsorption of 2 $NH_3$ molecules at $H_2$ pressure-$NH_3$ pressure 3 atm-$10^{-6}$ atm as shown in Figs. 3(c) and 3(d), respectively. This is ascribed to the Gibbs factor for the adsorption for 2 $NH_3$ molecules which dominates more than that for the adsorption for 5 $H_2$ molecules ($e^{2(\mu_{NH_3} - \varepsilon_{02})/kT} >> e^{5(\mu_{H_2} - \varepsilon_{50})/kT}$ where $\mu_{H_2} = -0.29$ eV, $\varepsilon_{50} = -0.38$ eV, $\mu_{NH_3} = -0.74$ eV, $\varepsilon_{02} = -1.18$ eV). This result shows that, without a change of temperature, $H_2$ molecules with a large binding energy of ~0.4−0.6 eV can be released by ammonia gas with a low pressure. From this result, we can suggest an approach to $H_2$ desorption on TM-decorated structures using the $H_2$-$NH_3$ partial pressure difference.

We consider the usable number of $H_2$ molecules per TM atom as a criterion for comparing different methods of the temperature variation and ammonia gas. To calculate the usable number of $H_2$ molecules at ambient conditions, we define the pressure-temperature adsorption conditions (50 atm-25 ºC) and the desorption conditions (3 atm-80 ºC) in the temperature variation method, and the hydrogen pressure-ammonia pressure-temperature adsorption conditions (50 atm-$10^{-9}$ atm-25 ºC) and the desorption conditions (3 atm-$10^{-6}$ atm-25 ºC) in the ammonia gas method. Under these conditions, the temperature and pressure reflect practical situations in vehicular operations. Then, $f$ at the adsorption conditions minus $f$ at the desorption conditions are the usable number of $H_2$ molecules per TM atom. The calculated usable number of $H_2$ molecules for all the structures with both methods is presented in Table I for comparison. For the Ti-decorated cPA, 4.80 $H_2$ molecules are usable, which is significantly increased when compared to 3.07 $H_2$ molecules with the temperature variation method. In contrast, the usable number of $H_2$ molecules for the Sc-decorated ETD is reduced to 0.36 compared to 2.18 with the temperature variation method because 0.38 $H_2$ and 0.97 $NH_3$ molecules adsorb on a Sc atom at the adsorption conditions by the Gibbs factor $e^{(\mu_{NH_3}-\varepsilon_{01})/kT}$ of the adsorption for one $NH_3$ molecule which dominates. For the case of the Sc-decorated cPA, the usable number of $H_2$ molecules is increased to 2.86 compared to 1.28 with the temperature variation method because 2.98 $H_2$ and 1.00 $NH_3$ molecules adsorb on a Sc atom at the adsorption conditions by $e^{(3\mu_{H_2}+\mu_{NH_3}-4\varepsilon_{31})/kT}$ which dominates ( $\mu_{H_2}=-0.21$ eV, $\mu_{NH_3}=-0.91$ eV, $\varepsilon_{31}=-0.49$ eV ). These results show that the ammonia gas method is effective for a system with a large binding energy of $H_2$ molecules, and is as efficient as the temperature variation method. Since no temperature increase is needed while used for vehicles, it may be more convenient in practical situations than the temperature variation method.

Next, we estimate the amount of ammonia needed to fully release stored hydrogen of 5 kg which is necessary to achieve a driving range of about 500 km.[26,27] To release 5 kg $H_2$ adsorbed on the Sc-, Ti-, and V-decorated cPA (ETD), the necessary amount of ammonia is 15 (15), 17 (28), and 15 (14) kg, respectively. For instance, in the case of V-decorated ETD, 14 kg (=5×17×0.87/(2.73×2) kg) $NH_3$ is obtained from the usable number of molecules per the used number of ammonia molecules (i.e., 2.73 $H_2$ per 0.87 $NH_3$) from the Table I. We find that the ratio of the usable number of $H_2$ molecules to the used number of $NH_3$ molecules should be more than 4 (i.e., $\overline{N}_{use}^{H_2}/\overline{N}_{use}^{NH_3} \geq 4$) to release 5 kg $H_2$ by less than 10 kg $NH_3$ which may be desirable for mobile applications.

Using Eq. (1) with energy $-\varepsilon_{0m}$, we evaluate the temperature and the pressure for desorbing ammonia molecules adsorbed on TM atoms after using the stored hydrogen. We find that, under a pressure of ~$10^{-6}$ atm, the $NH_3$ molecules on the Sc- and V-decorated ETD are desorbed at a temperature of ~145 $^o$C as shown in Figs. 4(a) and 4(b), respectively, because the chemical potential of $HN_3$ gas is lower than the binding energy at the conditions ($\mu_{NH_3} = -1.09$ eV, $\varepsilon_{01} = -0.99$ eV for the Sc-decorated ETD, and $\varepsilon_{01} = -1.02$ eV for the V-decorated ETD). The temperature and pressure are easily achievable if the releasing of ammonia from the storage tank is done off-board vehicles with mechanical pumps such as rotary or dry scroll pumps (accessible pressure ~$10^{-6}$ atm).

Next, we estimate the optimal binding energy of $H_2$-releasing molecule meeting the following requirements: (1) The usable number of $H_2$ molecules should be maximized by low partial pressure of $H_2$-releasing gas (~$10^{-6}$ atm) at room temperature and (2) the $H_2$-releasing molecules adsorbed on TM atoms should be released at feasible conditions of the temperature and the pressure (~125 $^o$C and ~$10^{-6}$ atm) after using the stored hydrogen. To meet the first requirement, the conditions of $e^{(\mu_{H_2}-\varepsilon_{10})/kT} >> e^{(\mu_\chi-\varepsilon_{01})/kT}$ at $H_2$ pressure- $\chi$ pressure-temperature of 50 atm-$10^{-9}$ atm-25 $^o$C and of $e^{(\mu_{H_2}-\varepsilon_{10})/kT} << e^{(\mu_\chi-\varepsilon_{01})/kT}$ at 3 atm-$10^{-6}$ atm-25 $^o$C should both be satisfied if it is assumed that one $H_2$ or $\chi$ molecule adsorbs on a TM atom. As a result, the binding energy should be 0.9 eV $< -\varepsilon_{01} < 1.3$ eV when the binding energy of $H_2$ molecule $-\varepsilon_{10}$ is approximated to be 0.5 eV. To meet the second requirement, the Gibbs factor for the binding of $H_2$-releasing molecule should be negligible ($e^{(\mu_g - \varepsilon_{01})/kT} \ll 1$, $-\varepsilon_{01} < 1.1$ eV) at ~125 °C and ~$10^{-6}$ atm. Therefore, it is estimated that the optimal binding energy is in the energy window of ~0.9−1.1 eV. This optimal binding energy may be revised as the binding energy of $H_2$ molecules and the adsorption and desorption conditions we chose.

We also examine different well-defined gases, nitrogen, ethylene, acetylene, oxygen, and water for a $H_2$-releasing molecule. Binding energies of $N_2$, $C_2H_4$, $C_2H_2$, $O_2$, and $H_2O$ molecules on TM atoms are calculated to be ~1.0, 1.5, 3.0, 7.0, and 3.0 eV, respectively. These molecules, except for $N_2$, are not suitable for a $H_2$-releasing molecule because of the optimal binding energy of the molecule (~1 eV). However, nitrogen gas might be not suitable because nitrogen and hydrogen on surfaces of transition metal materials combine to produce ammonia under high temperatures and very high pressures (the so-called Haber-Bosch process).[28] In contrast, ammonia gas is suitable for a $H_2$-releasing molecule because the binding energy of $NH_3$ molecule is ~1 eV, and no further chemical processes are expected on TM atoms. Furthermore, ammonia is very efficient and convenient for practical applications because it is quite light compared to the other well-defined gases and its phase is liquid (−33 °C boiling point).

Next, we consider poisoning effects of ammonia gas on fuel cell. According to a experimental paper[29], the poisoning effect of ammonia gas on the proton exchange membrane fuel cell (PEMFC) occurs at concentrations beyond 20 ppm. Since the concentration of $NH_3$ per $H_2$ of ~$10^{-3}$~$10^{-1}$ ppm in our working pressure range is much lower than 20 ppm, ammonia gas may do not affect the fuel cell performance. Therefore, some purification is not necessary and the partial pressure method we propose is practical.

## IV. CONCLUSION

In conclusion, we have found that, similar to the mechanism of CO poisoning of hemoglobin through the $O_2$-CO partial pressure difference, $H_2$ molecules adsorbed on transition metal atoms are released with the $H_2$-$NH_3$ partial pressure difference at room temperature. We feel that this suggestion represents a new approach to hydrogen desorption in nanostructured hydrogen-storage materials.

## ACKNOWLEDGMENTS

This research was supported by the Center for Nanotubes and Nanostructured Composites funded by the Korean Government MOST/KOSEF, and the Korean Government MOEHRD, Basic Research Fund No. KRF-2006-341-C000015. The work at Tsinghua was supported by the Ministry of Science and Technology of China (Grant Nos. 2006CB605105 and 2006CB0L0601), and the Natural Science Foundation of China and the Ministry of Education of China (Grant Nos. 10325415 and 10674077).

## REFERENCES

[1] G. J. Kubas, Science **314**, 1096 (2006).

[2] G. J. Kubas, J. Organomet. Chem. **635**, 37 (2001).

[3] Y. Zhao, Y.-H. Kim, A. C. Dillon, M. J. Heben, and S. B. Zhang, Phys. Rev. Lett. **94**, 155504 (2005).

[4] T. Yildirim and S. Ciraci, Phys. Rev. Lett. **94**, 175501 (2005).

[5] H. Lee, W. I. Choi, and J. Ihm, Phys. Rev. Lett. **97**, 056104 (2006).

[6] W. H. Shin, S. H. Yang, W. A. Goddard III, and J. K. Kang, Appl. Phys. Lett. **88**, 053111 (2006).

[7] E. Durgun, S. Ciraci, W. Zhou, and T. Yildirim, Phys. Rev. Lett. **97**, 226102 (2006).

[8] N. Park, S. Hong, G. Kim, and S.-H. Jhi, J .Am. Chem. Soc. **129**, 8999 (2007).

[9] X. Hu, B. O. Skadtchenko, M. Trudeau, and D. M. Antonelli, J. Am. Chem. Soc. **128**, 11740 (2006).

[10] A. Hamaed, M. Trudeau, and D. M. Antonelli, J. Am. Chem. Soc. **130**, 6992 (2008).

[11] A. B. Philips and B. S. Shivaram, Phys. Rev. Lett. **100**, 105505 (2008).

[12] G. Kim, S.-H. Jhi, N. Park, S. G. Louie, and M. L. Cohen, Phys. Rev. B **78,** 085408 (2008).

[13] S. Meng, E. Kaxiras, and Z. Zhang, Nano Lett. **7**, 663 (2007).

[14] Y. Zhao *et al.*, Nano Lett. **8**, 157 (2008).

[15] B. Kiran *et al.*, J. Chem. Phys. **124**, 224703 (2006).

[16] Y.-H. Kim, Y. Zhao, A. Williamson, M. J. Heben, and S. B. Zhang, Phys. Rev. Lett. **96**, 016102 (2006).

[17] R. Banerjee *et al.*, Nature **217**, 23 (1968).

[18] C. R. Johnson *et al.*, Biophys. Chem. **45**, 7 (1992).

[19] J. Ihm, A. Zunger, and M. L. Cohen, J. Phys. *C* **12**, 4409 (1979).

[20] J. P. Perdew and Y. Wang, Phys. Rev. B **45**, 13244 (1992).

[21] M. L. Cohen, M. Schlüter, J. R. Chelikowsky, and S. G. Louie, Phys. Rev. B **12**, 5575 (1975).

[22] M. C. Nguyen *et al.*, Solid State Comm. **147**, 419 (2008).

[23] L. C. Fernández-Torres *et al.*, Surf. Sci. **511**, 121 (2002).

[24] The experimental chemical potential of $H_2$ gas was used from *Handbook of Chemistry and Physics*, edited by D. R. Lide (CRC Press, New York, 1994), 75th ed. Since the values for the experimental chemical potential of $NH_3$ gas are not available in our pressure range, the chemical potential of ideal gas for $NH_3$ gas was used. This approximation may be reliable because of considerably low pressure range ($10^{-6} \sim 10^{-9}$ atm).

[25] http://www.eere.energy.gov/hydrogenandfuelcells/mypp/.

[26] L. Schlapbach and A. Züttel, Nature (London) **414**, 353 (2001).

[27] G. W. Crabtree, M. S. Dresselhaus, and M. V. Buchanan, Phys. Today, **57**, No. **12**, 39 (2004).

[28] J. M. Smith *et al.*, J. Am. Chem. Soc. **123**, 9222 (2001).

[29] N. Rajalakshmi, T. T. Jayanth, and K. S. Dhathathreyan, Fuel Cells, **3**, 177 (2003).

**Table I**. Comparison for the usable number of $H_2$ molecules obtained using the temperature variation and ammonia gas methods in TM-decorated cPA and ETD. Here $f_{H_2} \equiv N_{ads}^{H_2}$ is the number of adsorbed $H_2$'s per TM atom at the condition of adsorption (50 atm-25 °C), and $f_{H_2} \equiv N_{des}^{H_2}$ is the number of adsorbed $H_2$'s per TM atom at the condition of desorption (3 atm-80 °C) in the absence of ammonia gas. $f_{H_2} \equiv \overline{N}_{ads}^{H_2}$ ($f_{NH_3} \equiv \overline{N}_{ads}^{NH_3}$) is the number of adsorbed $H_2$ ($NH_3$)'s per TM atom at the conditions of adsorption ($p_{H_2} = 50\,\text{atm}$ - $p_{N_2} = 10^{-9}\,\text{atm}$ - $T = 25\,°C$), and $f_{H_2} \equiv \overline{N}_{des}^{H_2}$ ($f_{NH_3} \equiv \overline{N}_{des}^{NH_3}$) is the number of adsorbed $H_2$ ($NH_3$)'s per TM atom at the conditions of desorption ($p_{H_2} = 3\,\text{atm}$ - $p_{NH_3} = 10^{-6}\,\text{atm}$ - $T = 25\,°C$) in the presence of ammonia gas. The usable (used) number of $H_2$ ($NH_3$) molecules per TM atom is obtained from $N_{use}^{H_2} = N_{ads}^{H_2} - N_{des}^{H_2}$ or $\overline{N}_{use}^{H_2} = \overline{N}_{ads}^{H_2} - \overline{N}_{des}^{H_2}$ ($\overline{N}_{use}^{NH_3} = \overline{N}_{des}^{NH_3} - \overline{N}_{ads}^{NH_3}$).

| Materials | $N_{ads}^{H_2}$ | $N_{des}^{H_2}$ | $\overline{N}_{ads}^{H_2}$ | $\overline{N}_{des}^{H_2}$ | $\overline{N}_{ads}^{NH_3}$ | $\overline{N}_{des}^{NH_3}$ | $N_{use}^{H_2}$ | $\overline{N}_{use}^{H_2}$ | $\overline{N}_{use}^{NH_3}$ |
|---|---|---|---|---|---|---|---|---|---|
| **Sc-cPA** | 1.28 | 0.00 | 2.98 | 0.12 | 1.00 | 2.00 | 1.28 | 2.86 | 1.00 |
| **Ti-cPA** | 5.00 | 1.93 | 4.91 | 0.11 | 0.04 | 2.00 | 3.07 | 4.80 | 1.96 |
| **V-cPA** | 3.95 | 2.85 | 3.00 | 0.45 | 1.00 | 1.88 | 1.10 | 2.55 | 0.88 |
| **Sc-ETD** | 2.18 | 0.00 | 0.38 | 0.02 | 0.97 | 1.28 | 2.18 | 0.36 | 0.31 |
| **Ti-ETD** | 3.43 | 1.05 | 3.43 | 1.93 | 0.00 | 1.00 | 2.38 | 1.50 | 1.00 |
| **V-ETD** | 3.05 | 1.25 | 2.89 | 0.16 | 0.13 | 1.00 | 1.80 | 2.73 | 0.87 |

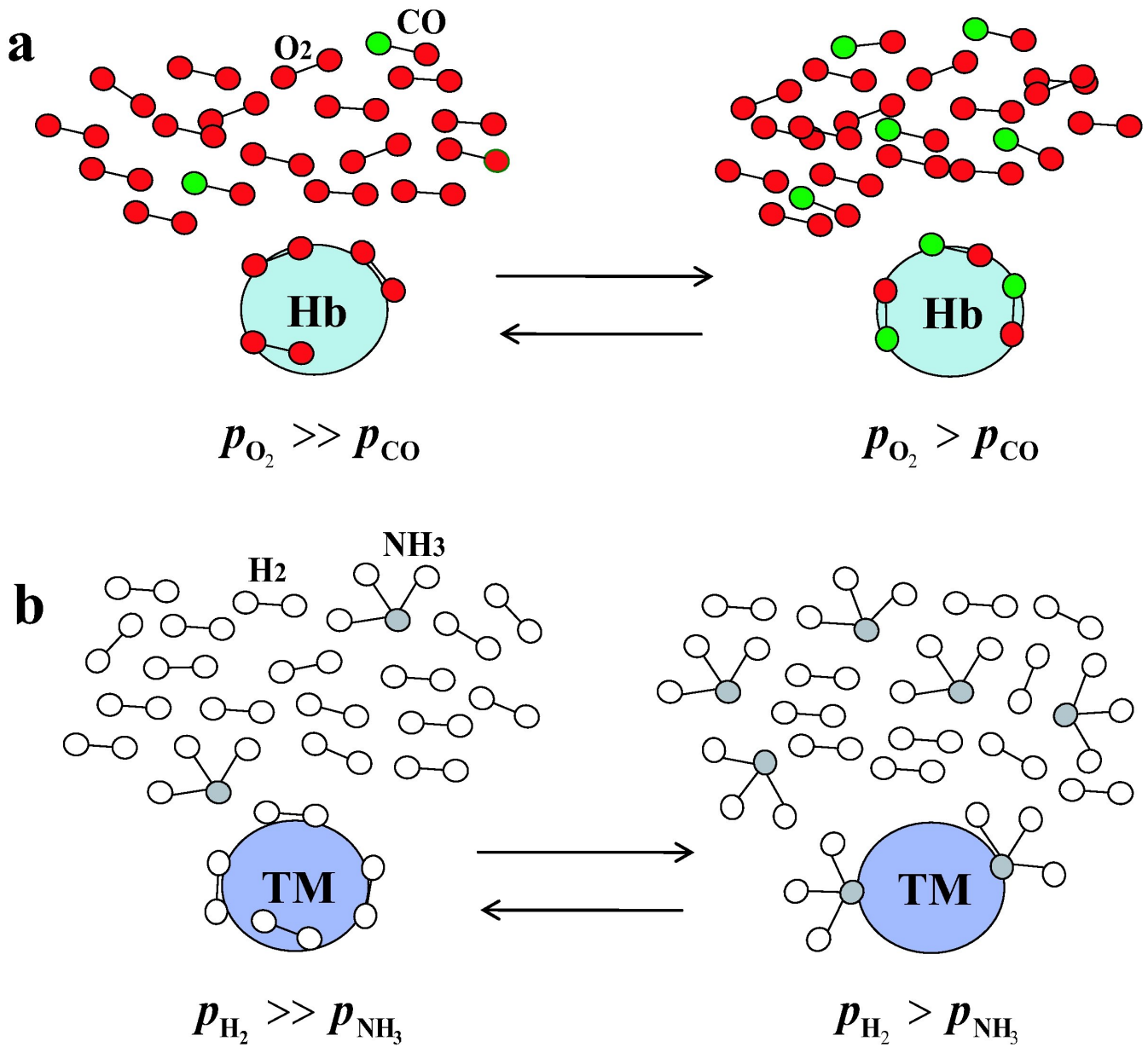

FIG. 1 (Color online). Schematic of the mechanism of releasing $H_2$ molecules with the partial pressure difference. (a) Illustration of the mechanism of carbon monoxide poisoning of hemoglobin with the oxygen-carbon monoxide partial pressure difference. In the condition of $p_{O_2} \gg p_{CO}$, $O_2$ molecules adsorb on hemoglobin. When the pressure of CO gas approaches that of $O_2$ gas $p_{O_2} \gtrsim p_{CO}$, the $O_2$ molecules are released by the adsorption of CO molecules. (b) Illustration of controlled adsorption of $H_2$ molecules on a TM atom with the hydrogen-ammonia partial pressure difference employing the mechanism of carbon monoxide poisoning.

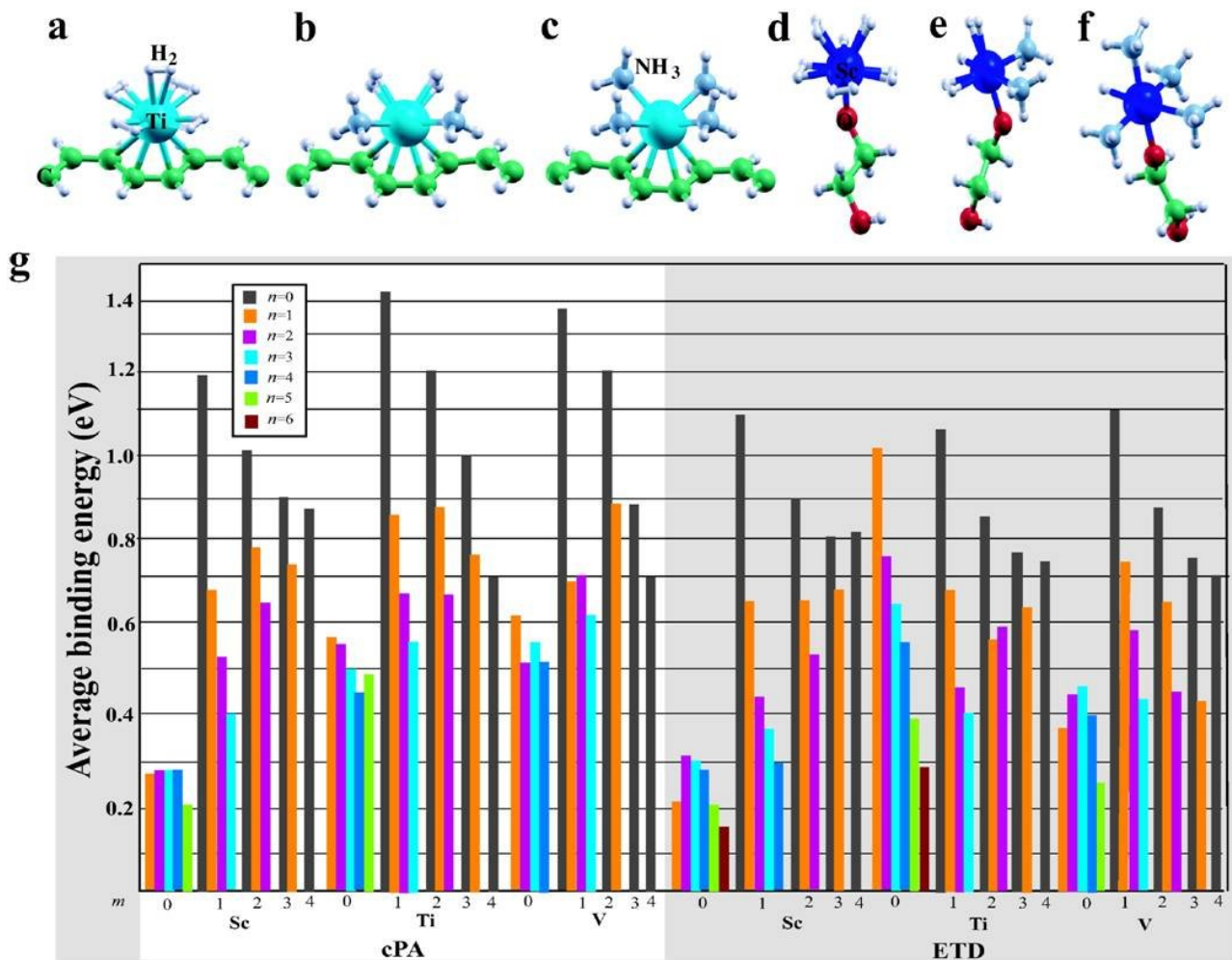

FIG. 2 (Color online). Optimized atomic geometries of TM-decorated cPA and ETD with $H_2$ and $NH_3$ molecules attached to a TM atom. (a)−(c) ((d)−(f)) show cPA (ETD) with five $H_2$ molecules, two $H_2$ and $NH_3$ molecules, and four $NH_3$ molecules attached per Ti (Sc) atom, respectively. (g) Calculated static binding energy (eV) per the number of adsorbed $H_2$ and $NH_3$ molecules on a TM atom attached to cPA and ETD as a function of the number of $H_2$ molecules ($n$) and $NH_3$ molecules ($m$).

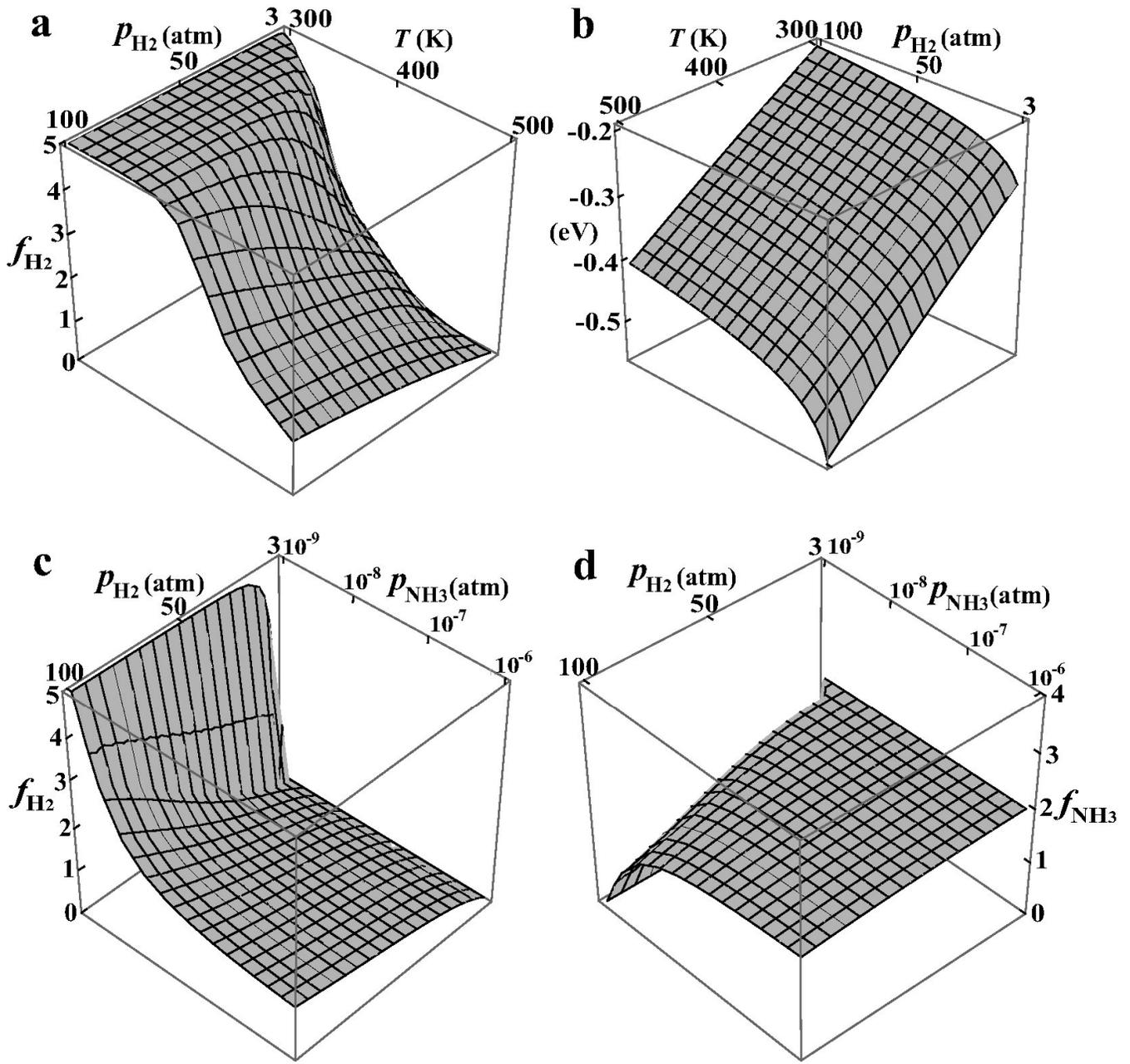

FIG. 3. Occupation number of $H_2$ and $NH_3$ molecules as hydrogen pressure-temperature ($f_{H_2}$ - $p_{H_2}$ -$T$ diagram) vs hydrogen pressure-ammonia pressure ($f_{H_2(NH_3)}$ - $p_{H_2}$ - $p_{NH_3}$ diagram). (a) Occupation number of $H_2$ molecules as a function of the pressure and temperature, ($f_{H_2}$ - $p_{H_2}$ -$T$ diagram), in Ti-decorated cPA. (b) The chemical potential of hydrogen gas as a function of the pressure and the temperature. (c) and (d) ($f_{H_2}$ - $p_{H_2}$ - $p_{NH_3}$ diagram) and ($f_{NH_3}$ - $p_{H_2}$ - $p_{NH_3}$ diagram)at 25 ºC in Ti-decorated cPA, respectively.

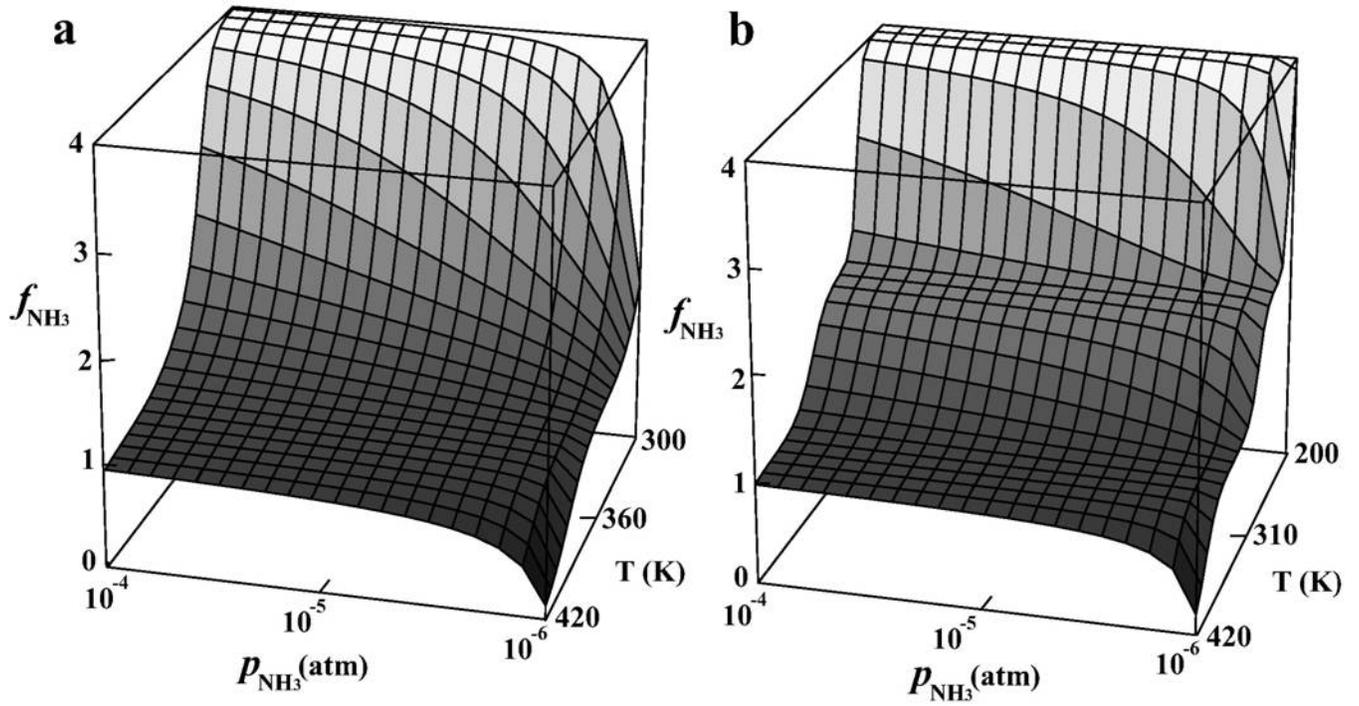

FIG. 4. Occupation number of NH₃ molecules as a function of the ammonia pressure and the temperature ($f_{NH_3}$ - $p_{NH_3}$ - $T$ diagram) (a) in Sc-decorated ETD and (b) in V-decorated ETD.